\newcommand{\version}{February 24, 2014}
\documentclass[a4paper,twoside,11pt,pdftex]{article}

\newcommand{\nofortschrphys}[1]{#1}
\newcommand{\fortschrphys}[1]{}

\fortschrphys{\usepackage{times,cite,w-thm,w-greek}}
\nofortschrphys{
\usepackage[bindingoffset=0.4cm,textheight=22.5cm,hdivide={2.7cm,*,2.75cm}, vdivide={*,22cm,*}]{geometry}
\usepackage{amsmath,amsfonts}
\usepackage[numbers,square,comma,sort&compress]{natbib}
}
\usepackage{amssymb}
\usepackage[pdftex]{graphicx}
\IfFileExists{dsfont.sty}
	{\usepackage{dsfont}
         \let\mathbb=\mathds
         \newcommand{\id}{\mathds{1}}}
	{\typeout{Package dsfont.sty was not found, using alternative macros.}
         \let\mathds=\mathbb
         \newcommand{\id}{\mbox{1 \kern-.59em {\rm l}}}}

\usepackage[pdftex,bookmarksnumbered=true,breaklinks=true]{hyperref}

\newcommand{\uim}{UV/IR mixing}
\newcommand{\nc}{non-commu\-ta\-tive}

\newcommand{\etal}{\textit{et al.}}

\newcommand{\starco}[2]{\left[ #1\stackrel{\star}{,}#2\right] }		
\newcommand{\pa}{\partial}						
\newcommand{\ri}{{\rm i}}						
\newcommand{\re}{{\rm e}}						
\renewcommand{\k}{\tilde{k}}						
\newcommand{\p}{\tilde{p}}						
\renewcommand{\d}{\delta}

\renewcommand{\th}{\theta}

\renewcommand{\l}{\lambda}
\newcommand{\m}{\mu}
\newcommand{\n}{\nu}

\newcommand{\inv}[1]{\frac{1}{#1}}				

\newcommand{\intk}{\int\! d^4k}					
\newcommand{\intx}{\int\! d^4x}						
\newcommand{\nn}{\nonumber}

\newcommand{\wsq}{\widetilde{\square}}
\newcommand{\ig}{{\rm i}g}

\nofortschrphys{
\title{\begin{flushright}
        {\small LA-UR-14-21128}\vspace*{2em}
       \end{flushright}
Gauge Fields on Non-Commutative Spaces \\and Renormalization
}
\date{\version}

\author{Daniel N. Blaschke}
}

\begin{document}
\fortschrphys{
\DOIsuffix{theDOIsuffix}
\Volume{55}
\Month{01}
\Year{2007}
\pagespan{1}{}
\Receiveddate{XXXX}
\Reviseddate{XXXX}
\Accepteddate{XXXX}
\Dateposted{XXXX}

\title[Gauge Fields on NC Spaces and Renormalization]{Gauge Fields on Non-Commutative Spaces \\and Renormalization}

\author[D.N. Blaschke]{Daniel N. Blaschke\inst{1,}%
  \footnote{Corresponding author\quad E-mail:~\textsf{dblaschke@lanl.gov},
            Phone: +1\,505\,667\,0849}}
\address[\inst{1}]{Los Alamos National Laboratory, Theory Division, Los Alamos, NM, 87545, USA}

}
\nofortschrphys{
\thispagestyle{empty}
\maketitle
\begin{center}
\renewcommand{\thefootnote}{\fnsymbol{footnote}}
\vspace{-0.3cm}Los Alamos National Laboratory, Theory Division\\Los Alamos, NM, 87545, USA\\[0.5cm]
\ttfamily{E-mail: dblaschke@lanl.gov}
\end{center}
\vspace{1.5em}
}

\begin{abstract}
Constructing renormalizable models on {\nc} spaces constitutes a big challenge.
Only few examples of renormalizable theories are known, such as the scalar Grosse-Wulkenhaar model.
Gauge fields are even more difficult, since new renormalization techniques are required which are compatible with the inherently non-local setting on the one hand, and also allow to properly treat the gauge symmetry on the other hand.
In this proceeding (which is based on my talk given at the ``Workshop on Noncommutative Field Theory and Gravity'' in Corfu/Greece, September 8 -- 15, 2013),
I focus on this last point and present new extensions to existing renormalization schemes (which are known to work for gauge field theories in commutative space) adapted to {\nc} Moyal space.
\end{abstract}

\fortschrphys{\maketitle}

\nofortschrphys{
\newpage
}
%
\section{Introduction}
\label{sec:intro}
%
In this proceeding we review the discussion of the BPHZ renormalization of scalar and gauge models in {\nc} 4-dimensional Euclidean space initiated in References~\cite{Blaschke:2012ex,Blaschke:2013cba}.

Quantum field theories on {\nc} spaces, in particular on Moyal space, have been studied for some time (for a review see~\cite{Szabo:2001,Rivasseau:2007a,Blaschke:2010kw} and references therein), and although problems preventing successful renormalization have been overcome in the scalar cases, such as the Grosse-Wulkenhaar model~\cite{Grosse:2003, Grosse:2004b} and the translation invariant $1/p^2$-model by Gurau \etal~\cite{Gurau:2009}, gauge theories have so far eluded us\cite{Blaschke:2009c}:
Even though some propositions how to restore renormalizability in those cases have been put forward, there exists no proof for their renormalizability.
Indeed, the methods considered so far for the quantization of scalar field theories on {\nc} space (such as Multiscale Analysis) cannot easily be applied to gauge field theories due to the fact that they break the gauge symmetry.
This is a strong motivation for trying to generalize the approach of BPHZ (Bogoliubov, Parasiuk, Hepp, Zimmermann) to the {\nc} setting, since it does not require the introduction of a regularization and has been proven to be a powerful tool for field theories with local symmetries on commutative space, see e.g. References~\cite{Piguet:1995,Schweda-book:1998}.

Before considering the BPHZ approach to the {\nc} setting, it is useful to recall the origin of {\uim} problem~\cite{Minwalla:1999px}:
In {\nc} space, the star product leads to the presence of phase factors in various (``non-planar'') Feynman graphs of field theories.
For large values of the internal momentum $k$, the rapid oscillations of the phase factor have a regularizing effect upon integration over $k$ leading to a finite result for any $\tilde p \neq 0$ (where $p$ is the external momentum).
However, there are additional contributions of the same form as in commutative space which are independent of such phase factors and hence UV-divergent.
In addition, the non-planar integrals are singular for small values of the external momentum (i.e. IR-divergent)~\cite{Minwalla:1999px,Blaschke:2008a}.
These new types of singularities destroy renormalizability unless additional terms are included in the action as has been done successfully in the scalar case in Refs.~\cite{Grosse:2003, Grosse:2004b,Gurau:2009}.

In the following section, we illustrate the main ideas behind the generalization of the BPHZ procedure to the {\nc} setting using a scalar $\phi^{\star4}$.
We briefly discuss the results of~\cite{Blaschke:2012ex,Blaschke:2013cba} and describe how to apply the new method to a candidate for a renormalizable gauge field model on {\nc} space out forward in~\cite{Blaschke:2009e,Blaschke:2010ck}.

\section{BPHZ applied to {\nc} \texorpdfstring{$\phi ^{\star4}$}{phi4}-theory}\label{sec:nc-scalar-theory}
The model under consideration is defined  at the classical level  by the  action (see e.g. Ref.~\cite{Szabo:2001})
\begin{align}
\label{action}
\Gamma^{(0)}[\phi]=
\frac 12 \int d^4x \left( \partial^\mu \phi \star \partial_\mu \phi+m^2 \, \phi\star \phi
 \right)
 \, + \,
\frac \l{4!} \int d^4x \left(\phi\star\phi\star\phi\star\phi \right)
\,,
\end{align}
where the Moyal star product is defined as
$
\left( f \star g \right) (x):=  \re ^{\frac{\ri}{2}\th^{\m\n}\pa^x_\m\pa^y_\n}f(x)g(y) \big|_{x=y}
$ with $
\th^{\m\n} = - \th^{\n\m}
$ constant.
Introducing the Fourier components
 $\tilde\phi(k)$  of $\phi$
we find that the propagator in momentum space is given by
$
\tilde\Delta (k)=({k^2+m^2})^{-1}
$,
and that the interaction term can be expressed
in terms of the variables $\tilde k_\mu \equiv \theta_{\mu\nu}k^\nu$ by
\begin{align}
\nonumber
\Gamma^{(0)}_{\text{int}}[\phi]=\frac{1}{4!} \intk_1\ldots d^4k_4 \, \tilde\phi(k_1)\tilde\phi(k_2)\tilde\phi(k_3)\tilde\phi(k_4) \, (2\pi)^4 \, \delta^{(4)}(k_1+k_2+k_3+k_4)
\, \bar{\lambda} \, ,
\end{align}
with
\begin{align}
\bar\lambda \equiv \frac{\lambda}{3}\left[\cos\!\left(\tfrac{k_1\k_2}{2}\right)\cos\!\left(\tfrac{k_3\k_4}2\right)
+\cos\!\left(\tfrac{k_1\k_3}2\right)\cos\!\left(\tfrac{k_2\k_4}2\right)
+\cos\!\left(\tfrac{k_1\k_4}2\right)\cos\!\left(\tfrac{k_2\k_3}2\right)\right]
\,. \label{eq:nc-scalar-coupling}
\end{align}
Hence, in comparison to the commutative $\phi^4$-theory, the interaction vertex of the {\nc} $\phi^{\star4}$-theory is characterized by a modified coupling in momentum space ($\lambda$ becomes $\bar{\lambda}$).

The quantization of this model and the renormalization of related scalar field models has been discussed over the last fifteen years, see for instance reference~\cite{Blaschke:2012ex} for a brief review and list of references.
In the latter work it was pointed out that the usual BPHZ momentum space subtraction scheme (which consists of subtracting appropriate polynomials in the external momentum from the integrand of divergent integrals) cannot be applied in {\nc}  theories, e.g. for a (non-planar) integral of the form
\begin{align}
J (p) \equiv
 \intk \,  \frac{\cos(k\p)}{[(p+k)^2+m^2][k^2+m^2]}
 \,. \label{eq:4pfct}
\end{align}
The problem is due to the phase factor $\cos(k\p)$ which is at the origin of the UV/IR mixing problem, i.e. the appearance of an IR-singularity for small values of the external momentum $p$.
Therefore, a modified subtraction scheme was proposed in Ref.~\cite{Blaschke:2012ex}:
It consists of considering $p$ and $\p$ as independent variables (though satisfying $p \tilde p =0$) when performing the subtraction:
In particular, one subtracts from the integrand its Taylor series expansion with respect to the external momentum $p$ around $p=0$ up to the order of divergence of the graph, while maintaining the phase factors.
Thus, for these non-planar graphs, our modified BPHZ subtraction amounts to an IR-subtraction rather than a UV-subtraction.
Thus, for the integral \eqref{eq:4pfct}  one considers
\begin{align}
J^{{\rm finite}} (p)
\equiv
\intk\left( \frac{\cos(k\p)}{[(p+k)^2+m^2][k^2+m^2]} - \frac{\cos(k\p)}{[k^2+m^2]^2} \right)
 \,. \label{eq:checkJ}
\end{align}
For the subtracted non-planar one-loop $2$-point function of the model \eqref{action}, we hence get a vanishing result (as one also does for the planar, UV-divergent diagram by virtue of the standard BPHZ subtraction scheme).
For the subtracted non-planar one-loop $4$-point
function \eqref{eq:checkJ}, one obtains a result which is regular in $\tilde p$, as shown in~\cite{Blaschke:2012ex}.

By proceeding in this way, the one-loop renormalization of the theory could be carried out for the (na\"ive) $\phi^{\star4}$-theory described by the action \eqref{action}, as well as for this action supplemented by a $1/p^2$-term which is known to overcome the UV/IR mixing problem while maintaining the translation invariance of the model~\cite{Gurau:2009}.
In Ref.~\cite{Blaschke:2013cba} this modified BPHZ subtraction scheme was then successfully applied to the scalar sunrise graph as an example for a two-loop graph with overlapping divergences (applying also the forest formula of Zimmermann~\cite{Zimmermann:1969} in the {\nc} setting).


Concerning the additional $1/p^2$-term in the model~\cite{Gurau:2009}, consider the following:
In UV-divergent planar diagrams, the cut-off regularization exhibits the degree (power of $\Lambda$) of the UV-divergence which determines also the degree of the polynomial in $p$ which in turn is considered for the standard BPHZ subtraction.
The ambiguity involved in the standard BPHZ subtraction (corresponding to a finite renormalization) is a polynomial in $p$ whose order is the superficial degree of UV-divergence of the diagram under consideration.

A non-planar diagram and the regularized version of the corresponding planar diagram have the same form up to the replacement $\Lambda ^2 \leadsto 4/{\p}^{2}$.
Hence, one expects that the ambiguity involved in the modified BPHZ subtraction amounts to a polynomial in $1/\p^{2}$ whose degree is determined by the degree of the IR-singularity of the non-planar graph.
In particular, for the expansion around $p=0$ for the modified BPHZ subtraction, the ambiguity is a polynomial in $p$ (with coefficients depending on the parameter $\p^{2}$ which is considered as an independent variable), the degree of this polynomial coinciding
with the degree of the IR-singularity of the non-planar graph.
All coefficients of this polynomial must, of course, have the correct dimension.
For the non-planar tadpole graph, which has a quadratic IR-singularity, we thus get a term $A \, \phi ^2$ (with $A$ having the dimension of a mass squared) and a term  $(\pa^{\mu} \phi)(\pa_{\mu} \phi)$ --- but now there is a further possibility involving $\tilde p ^{\, 2}$.
Since $\theta^{\mu\nu}$ (parameterizing {\nc} space) has the dimension of length squared (${[\hat{x} ^{\mu} , \hat{x} ^{\nu}] } = \ri \theta^{\mu \nu} \id$), an extra term appears as ambiguity for the subtraction: $\tilde{\phi} \, ({\tilde p}^{\, 2})^{-1} \tilde{\phi}$ (or in configuration space $\phi\,\wsq^{-1}\phi$ with $
\wsq\equiv\widetilde{\pa}^\m\widetilde{\pa}_\m=\th^{\m\m'}\th_{\m\n'}\pa_{\m'}\pa^{\n'}
$).
Such a non-local term is admissible in a translation invariant scalar field theory on {\nc} space and must be included if it is not present in the initial Lagrangian, hence explaining the additional term in the action of~\cite{Gurau:2009}.
In fact~\cite{Blaschke:2012ex}, it is the only non-local counterterm which can appear in a translation invariant {\nc} scalar field model.

Hence, after including this term into the Lagrangian, the propagator  for the $\phi^{\star4}$-theory reads
\begin{align}
G(k)= & \;
\Big(k^2+m^2+\tfrac{a^2}{k^2}\Big)^{-1}
\,, \label{eq:propagator}
\end{align}
and has a ``damping'' behaviour for vanishing momentum~\cite{Gurau:2009},
$ \lim\limits_{k \to 0}G(k)=0$,
which allows to overcome  potential IR-divergences in higher loop graphs, i.e. it is in fact crucial that this term is included in the action in order to achieve renormalizability (if translation invariance is to be maintained).
In fact, the IR-divergence of the non-planar tadpole graph
becomes potentially problematic when this graph is inserted into a higher loop diagram, since the external momentum of the insertion then becomes the internal momentum $k$ over which one integrates:
The divergence for $k \to 0$ then represents a potential problem for the renormalizability.
However, the damping behaviour allows to overcome this problem~\cite{Gurau:2009,Blaschke:2008a} and indeed it has been proven to provide a renormalizable model~\cite{Gurau:2009}.


\section{Non-commutative gauge field theories}\label{ncgt}

One of the motivations for generalizing the BPHZ approach to the {\nc} setting is to develop a tool for the renormalization of {\nc} gauge theories since the usual approaches such as Multiscale Analysis break gauge invariance, e.g. see reference~\cite{Rivasseau:2007a} for a review.
In the following, we briefly describe how the modified BPHZ method applies to gauge theories, for simplicity $U_\star(1)$.

The ``na\"ive'' gauge field action on {\nc} Euclidean space is given by
\begin{align}
 S_{\textrm{YM}} [A] &= \inv4
  \intx \, F_{\m\n}\star F^{\m\n}\,,
  \quad {\rm with} \ \;
 F_{\m\n} \equiv
 \pa_\m A_\n-\pa_\n A_\m-\ig\starco{A_\m}{A_\n}
 \,, \label{eq:gaugefield-naiv}
\end{align}
and again exhibits {\uim}.
Hence it is non-renormalizable (see e.g.~\cite{Blaschke:2010kw} and references therein) unless the action is modified.
Inspired by the results achieved for the scalar models, various approaches have been proposed in recent years~\cite{Grosse:2007,Wallet:2007c,Blaschke:2007b,Blaschke:2008a,Vilar:2009,Blaschke:2009e,Blaschke:2010ck} --- see
also the discussion~\cite{Blaschke:2013gha} and references therein.
However, so far none of these models could be proven to be renormalizable, in part due to the lack of a renormalization scheme which is compatible with both non-commutativity and gauge symmetry~\cite{Blaschke:2009c}.

Let us take a closer look at the one-loop vacuum polarization
with Feynman gauge fixing.
Three Feynman graphs contribute~\cite{Blaschke:2013gha} and their sum
features a phase independent (UV divergent) part as well as a phase dependent (convergent) part.
The former is superficially quadratically
UV-divergent by power counting, however it is well known that gauge symmetry (i.e. the Ward identity $p^\m \Pi_{\m\n}=0$) reduces this degree of divergence to a logarithmic one.
On the other hand, the phase dependent contribution is UV-finite due to the regularizing effect of the cosine, but it develops a quadratic IR-singularity for $\p\to 0$, i.e.:
\begin{align}
 \Pi_{\m\n}
 &=\frac{2g^2}{\pi^2}\frac{\p_\m \p_\n}{(\p^{2})^2}
 \qquad
 \textrm{for} \quad \p^{2} \ll 1
 \, . \label{eq:gauge-IR-div}
\end{align}
This IR-divergence remains a quadratic one since it is compatible with the Ward identity $p^\m \Pi_{\m\n}=0$ following from the gauge symmetry due to the fact that $p\p=0$.
Furthermore, the one-loop correction to the $3A$-vertex yields a linearly infrared divergent term which is connected to the present quadratic IR-divergence via Slavnov-Taylor identities~\cite{Blaschke:2013gha}

In general, massless theories require additional regularization in the infrared regime.
However, such a regularization is potentially problematic for gauge models since a regulator mass generically violates gauge invariance\footnote{Unless it is implemented via a BRST-doublet, as has more recently been done in Ref.~\cite{Quadri:2003pq}.} --- see references~\cite{Blaschke:2013cba,Lowenstein:1975pd,Grassi:1995wr}.
In the commutative case, this issue is usually addressed
by using dimensional regularization.
However, this method is not appropriate in the {\nc} setting, in particular due to the {\uim}.
Furthermore, the IR-divergences of the type \eqref{eq:gauge-IR-div} arise from the UV-divergences via {\uim} and are at the origin of the non-renormalizability of the (na\"ive) gauge field model~\eqref{eq:gaugefield-naiv}.
Therefore, we will consider a gauge field model with additional terms in the action~\cite{Blaschke:2009e} which provide a damping in the infrared regime for the gauge field propagator similar to the one for the scalar $1/p^2$ model of Gurau \etal~\cite{Gurau:2009}.
Thus, the one-loop vacuum polarization in a Feynman-like gauge fixing becomes
$\Pi^{(a)}_{\m\n}(p) \equiv  \intk\, I^{(a)}_{\mu \nu}(p,\p,k)$
with\footnote{For simplicity, we neglect an extra non-local counterterm for the singularity \eqref{eq:gauge-IR-div}, thus setting the parameter $\sigma\!=\!0$ appearing in the  gauge field propagator of Ref.~\cite{Blaschke:2009e}.
Although for the present illustration we consider a Feynman-like gauge fixing with an additional damping factor in order to arrive at the simplest form of the gauge field propagator,
we note that the full model of Ref.~\cite{Blaschke:2009e} is based on the Landau gauge fixing (but may be generalized to other gauges along the lines of~\cite{Lavrov:2013boa}).}
\begin{align}
 I^{(a)}_{\m\n}(p,\p,k)& \equiv \frac{2g^2\left(1-\cos(k\p) \right)\left(4k_\m k_\n-3p_\m p_\n+2\d_{\m\n} (p^2-k^2)\right)}{(2\pi)^4\left[k^2+\frac{a^2}{k^2}\right]\left[(k+p)^2+\frac{a^2}{(k+p)^2}\right]}
 \,.
\end{align}
Applying the (modified) BPHZ scheme described in the previous section,
we find that~\cite{Blaschke:2013cba}
\begin{align}
  &\Pi_{\m\n}^{(a) \textrm{finite}}(p) \equiv  \intk \left(1-t_{p}^2\right)
  I^{(a)}_{\mu \nu}
   (p,\p,k) \nn \\
   &={2g^2}\!\int\!\frac{d^4k}{(2\pi)^4}
 \, \frac{\left(1\!-\!\cos(k\p)\right)}{N} \,
   \Bigg\{
   \frac{4 k_\m k_\n-2 \d_{\m\n} k^2}{N^2}
 \Bigg[p^2-\frac{a^2p^2}{\left(k^2\right)^2}+\frac{4(kp)^2}{N}  \left(\frac{3a^2}{(k^2)^2}\!-\!1\right) \!\!\Bigg]\nn\\
 &\qquad \qquad +\left[4k_\m k_\n-3p_\m p_\n+2\d_{\m\n}(p^2-k^2)\right] \Bigg[\inv{(k\!+\!p)^2+\frac{a^2}{(k+p)^2}}-\inv{N}\!\Bigg] \Bigg\}
   \,, \label{eq:intpun}
\end{align}
where the operator $t_{p}^2$ denotes a second order Taylor expansion  with respect to $p$ around $p=0$ (but keeping $\p\neq0$ and independent), and we introduced the abbreviation $N:=\left(k^2+\frac{a^2}{k^2}\right)$.
The integral \eqref{eq:intpun} may eventually be carried out further by using the decomposition~\cite{Blaschke:2008b,Blaschke:2009a}
$
\left( k^2+\frac{a^2}{k^2} \right)^{-1}= \inv2\sum_{\zeta=\pm1}\frac{1}{k^2+\ri a\zeta}
$.
However, the main point is that expression \eqref{eq:intpun} represents a UV- and IR-finite result.

\section{Conclusion}
\label{sec:con}

We have reviewed the modified BPHZ scheme put forward in References~\cite{Blaschke:2012ex,Blaschke:2013cba}, where it was argued that this method works for higher loop graphs involving overlapping divergences and that its application is unambiguous in the {\nc} setting.
In the scalar case this scheme implies the introduction of a non-local $1/p^2$-term into the action --- which is precisely the one allowed (and induced) by the star product, and the resulting action has previously been shown to define a renormalizable theory by application of Multiscale Analysis~\cite{Gurau:2009}.

Furthermore, we have pointed out that the application of the modified BPHZ scheme to {\nc} gauge field theories looks promising, although further investigations are required.

\fortschrphys{\begin{acknowledgement}}
\nofortschrphys{\subsection*{Acknowledgements}}
D.N. Blaschke is a recipient of an APART fellowship of the Austrian Academy of Sciences, and is also grateful for the hospitality of the theory division of LANL and its partial financial support.
\fortschrphys{\end{acknowledgement}}


\bibliographystyle{./../../custom1}
\bibliography{./../../articles,./../../books}

\begin{thebibliography}{10}
\expandafter\ifx\csname url\endcsname\relax
  \def\url#1{\texttt{#1}}\fi
\expandafter\ifx\csname urlprefix\endcsname\relax\def\urlprefix{\\URL }\fi
\providecommand{\eprint}[2][]{\url{#2}}
\small\itemsep=3pt
\tolerance 1414
\hbadness 1414
\emergencystretch 1.5em
\hfuzz 0.3pt
\widowpenalty=10000
\vfuzz \hfuzz
\raggedbottom

\bibitem{Blaschke:2012ex}
D.~N. Blaschke, T.~Garschall, F.~Gieres, F.~Heindl, M.~Schweda and
  M.~Wohlgenannt, \textit{{On the Renormalization of Non-Commutative Field
  Theories}}, \textit{Eur. Phys. J.} \textbf{C73} (2013) 2262,
  \href{http://www.arxiv.org/abs/1207.5494}{\texttt{[arXiv:1207.5494]}}.

\bibitem{Blaschke:2013cba}
D.~N. Blaschke, F.~Gieres, F.~Heindl, M.~Schweda and M.~Wohlgenannt,
  \textit{{BPHZ renormalization and its application to non-commutative field
  theory}}, \textit{Eur. Phys. J.} \textbf{C73} (2013) 2566,
  \href{http://www.arxiv.org/abs/1307.4650}{\texttt{[arXiv:1307.4650]}}.

\bibitem{Szabo:2001}
R.~J. Szabo, \textit{Quantum field theory on noncommutative spaces},
  \textit{Phys. Rept.} \textbf{378} (2003) 207--299,
  \href{http://www.arxiv.org/abs/hep-th/0109162}{\texttt{[arXiv:hep-th/0109162%
]}}.

\bibitem{Rivasseau:2007a}
V.~Rivasseau, \textit{Non-commutative renormalization,} {in
  \textit{S\'{e}minaire Poincar\'{e} X (2007) --- Espaces Quantiques}, B.
  {Duplantier} and {V. Rivasseau} eds., Birkh\"{a}user Verlag},
  \href{http://www.arxiv.org/abs/0705.0705}{\texttt{[arXiv:0705.0705]}}.

\bibitem{Blaschke:2010kw}
D.~N. Blaschke, E.~Kronberger, R.~I.~P. Sedmik and M.~Wohlgenannt,
  \textit{{Gauge Theories on Deformed Spaces}}, \textit{SIGMA} \textbf{6}
  (2010) 062,
  \href{http://www.arxiv.org/abs/1004.2127}{\texttt{[arXiv:1004.2127]}}.

\bibitem{Grosse:2003}
H.~Grosse and R.~Wulkenhaar, \textit{Renormalisation of {$\phi^4$} theory on
  noncommutative {$\mathbb R^2$} in the matrix base}, \textit{JHEP} \textbf{12}
  (2003) 019,
  \href{http://www.arxiv.org/abs/hep-th/0307017}{\texttt{[arXiv:hep-th/0307017%
]}}.

\bibitem{Grosse:2004b}
H.~Grosse and R.~Wulkenhaar, \textit{Renormalisation of {$\phi^4$} theory on
  noncommutative {$\mathbb R^4$} in the matrix base}, \textit{Commun. Math.
  Phys.} \textbf{256} (2005) 305--374,
  \href{http://www.arxiv.org/abs/hep-th/0401128}{\texttt{[arXiv:hep-th/0401128%
]}}.

\bibitem{Gurau:2009}
R.~Gurau, J.~Magnen, V.~Rivasseau and A.~Tanasa, \textit{A
  translation-invariant renormalizable non-commutative scalar model},
  \textit{Commun. Math. Phys.} \textbf{287} (2009) 275--290,
  \href{http://www.arxiv.org/abs/0802.0791}{\texttt{[arXiv:0802.0791]}}.

\bibitem{Blaschke:2009c}
D.~N. Blaschke, E.~Kronberger, A.~Rofner, M.~Schweda, R.~I.~P. Sedmik and
  M.~Wohlgenannt, \textit{{On the Problem of Renormalizability in
  Non-Commutative Gauge Field Models --- A Critical Review}}, \textit{Fortschr.
  Phys.} \textbf{58} (2010) 364,
  \href{http://www.arxiv.org/abs/0908.0467}{\texttt{[arXiv:0908.0467]}}.

\bibitem{Piguet:1995}
O.~Piguet and S.~P. Sorella, \textit{Algebraic renormalization: Perturbative
  renormalization, symmetries and anomalies}, \textit{Lect. Notes Phys.}
  \textbf{M28} (1995) 1--134.

\bibitem{Schweda-book:1998}
A.~Boresch, S.~Emery, O.~Moritsch, M.~Schweda, T.~Sommer and H.~Zerrouki,
  \textit{Applications of Noncovariant Gauges in the Algebraic Renormalization
  Procedure}, Singapore: World Scientific, 1998.

\bibitem{Minwalla:1999px}
S.~Minwalla, M.~Van~Raamsdonk and N.~Seiberg, \textit{Noncommutative
  perturbative dynamics}, \textit{JHEP} \textbf{02} (2000) 020,
  \href{http://www.arxiv.org/abs/hep-th/9912072}{\texttt{[arXiv:hep-th/9912072%
]}}.

\bibitem{Blaschke:2008a}
D.~N. Blaschke, F.~Gieres, E.~Kronberger, M.~Schweda and M.~Wohlgenannt,
  \textit{{Translation-invariant models for non-commutative gauge fields}},
  \textit{J. Phys. A: Math. Theor.} \textbf{41} (2008) 252002,
  \href{http://www.arxiv.org/abs/0804.1914}{\texttt{[arXiv:0804.1914]}}.

\bibitem{Blaschke:2009e}
D.~N. Blaschke, A.~Rofner, R.~I.~P. Sedmik and M.~Wohlgenannt, \textit{{On
  Non-Commutative $U_\star(1)$ Gauge Models and Renormalizability}}, \textit{J.
  Phys. A: Math. Theor.} \textbf{43} (2010) 425401,
  \href{http://www.arxiv.org/abs/0912.2634}{\texttt{[arXiv:0912.2634]}}.

\bibitem{Blaschke:2010ck}
D.~N. Blaschke, \textit{{A New Approach to Non-Commutative $U_\star(N)$ Gauge
  Fields}}, \textit{EPL} \textbf{91} (2010) 11001,
  \href{http://www.arxiv.org/abs/1005.1578}{\texttt{[arXiv:1005.1578]}}.

\bibitem{Zimmermann:1969}
W.~Zimmermann, \textit{{Convergence of Bogoliubov's Method of Renormalization
  in Momentum Space}}, \textit{Comm. Math. Phys.} \textbf{15} (1969) 208--234.

\bibitem{Grosse:2007}
H.~Grosse and M.~Wohlgenannt, \textit{Induced gauge theory on a noncommutative
  space}, \textit{Eur. Phys. J.} \textbf{C52} (2007) 435--450,
  \href{http://www.arxiv.org/abs/hep-th/0703169}{\texttt{[arXiv:hep-th/0703169%
]}}.

\bibitem{Wallet:2007c}
A.~de~Goursac, J.-C. Wallet and R.~Wulkenhaar, \textit{Noncommutative induced
  gauge theory}, \textit{Eur. Phys. J.} \textbf{C51} (2007) 977--987,
  \href{http://www.arxiv.org/abs/hep-th/0703075}{\texttt{[arXiv:hep-th/0703075%
]}}.

\bibitem{Blaschke:2007b}
D.~N. Blaschke, H.~Grosse and M.~Schweda, \textit{{Non-Commutative $U(1)$ Gauge
  Theory on $\mathbb{R}^4$ with Oscillator Term and BRST Symmetry}},
  \textit{Europhys. Lett.} \textbf{79} (2007) 61002,
  \href{http://www.arxiv.org/abs/0705.4205}{\texttt{[arXiv:0705.4205]}}.

\bibitem{Vilar:2009}
L.~C.~Q. Vilar, O.~S. Ventura, D.~G. Tedesco and V.~E.~R. Lemes, \textit{{On
  the Renormalizability of Noncommutative $U(1)$ Gauge Theory --- an Algebraic
  Approach}}, \textit{J. Phys. A: Math. Theor.} \textbf{43} (2010) 135401,
  \href{http://www.arxiv.org/abs/0902.2956}{\texttt{[arXiv:0902.2956]}}.

\bibitem{Blaschke:2013gha}
D.~N. Blaschke, H.~Grosse and J.-C. Wallet, \textit{{Slavnov-Taylor identities,
  non-commutative gauge theories and infrared divergences}}, \textit{JHEP}
  \textbf{06} (2013) 038,
  \href{http://www.arxiv.org/abs/1302.2903}{\texttt{[arXiv:1302.2903]}}.

\bibitem{Quadri:2003pq}
A.~Quadri, \textit{{Higher order nonsymmetric counterterms in pure Yang-Mills
  theory}}, \textit{J. Phys. G: Nucl. Part. Phys.} \textbf{30} (2004) 677,
  \href{http://www.arxiv.org/abs/hep-th/0309133}{\texttt{[arXiv:hep-th/0309133%
]}}.

\bibitem{Lowenstein:1975pd}
J.~H. Lowenstein, \textit{{Auxiliary Mass Formulation of the Pure Yang-Mills
  Model}}, \textit{Nucl. Phys.} \textbf{B96} (1975) 189.

\bibitem{Grassi:1995wr}
P.~A. Grassi, \textit{{Stability and renormalization of Yang-Mills theory with
  background field method: A Regularization independent proof}}, \textit{Nucl.
  Phys.} \textbf{B462} (1996) 524--550,
  \href{http://www.arxiv.org/abs/hep-th/9505101}{\texttt{[arXiv: hep-th/9505101%
]}}.

\bibitem{Lavrov:2013boa}
P.~M. Lavrov and O.~Lechtenfeld, \textit{{Gribov horizon beyond the Landau
  gauge}}, \textit{Phys. Lett.} \textbf{B725} (2013) 386--388,
  \href{http://www.arxiv.org/abs/1305.2931}{\texttt{[arXiv:1305.2931]}}.

\bibitem{Blaschke:2008b}
D.~N. Blaschke, F.~Gieres, E.~Kronberger, T.~Reis, M.~Schweda and R.~I.~P.
  Sedmik, \textit{Quantum Corrections for Translation-Invariant Renormalizable
  Non-Commutative $\Phi^4$ Theory}, \textit{JHEP} \textbf{11} (2008) 074,
  \href{http://www.arxiv.org/abs/0807.3270}{\texttt{[arXiv:0807.3270]}}.

\bibitem{Blaschke:2009a}
D.~N. Blaschke, A.~Rofner, M.~Schweda and R.~I.~P. Sedmik, \textit{{One-Loop
  Calculations for a Translation Invariant Non-Commutative Gauge Model}},
  \textit{Eur. Phys. J.} \textbf{C62} (2009) 433,
  \href{http://www.arxiv.org/abs/0901.1681}{\texttt{[arXiv:0901.1681]}}.

\end{thebibliography}

\end{document}